\documentclass[fleqn,10pt]{wlscirep}

\usepackage{amsfonts}
\usepackage{amsmath}
\usepackage{amssymb}
\usepackage{revsymb}
\usepackage{mathrsfs}

\usepackage{graphicx}
\usepackage{dcolumn}
\usepackage{bm}

\usepackage{longtable}
\usepackage{multirow, bigdelim}
\usepackage{array}

\usepackage{color}

\title{Laser-pulse-shape control of seeded QED cascades}

\author[1,*]{Matteo~Tamburini}
\author[1]{Antonino~Di~Piazza}
\author[1]{Christoph~H.~Keitel}
\affil[1]{Max-Planck-Institut f\"ur Kernphysik, Saupfercheckweg 1, D-69117 Heidelberg, Germany}

\affil[*]{matteo.tamburini@mpi-hd.mpg.de}

\begin{abstract}
QED cascades are complex avalanche processes of hard photon emission and electron-positron pair creation driven by ultra-strong electromagnetic fields. They play a fundamental role in astrophysical environments such as a pulsars' magnetosphere, rendering an earth-based implementation with intense lasers attractive. In the literature, QED cascades were also predicted to limit the attainable intensity in a set-up of colliding laser beams in a tenuous gas such as the residual gas of a vacuum chamber, therefore severely hindering experiments at extreme field intensities. 
Here, we demonstrate that the onset of QED cascades may be either prevented even at intensities around $10^{26}\text{ W/cm$^{2}$}$ with tightly focused laser pulses and low-$Z$ gases, or facilitated at intensities below $10^{24}\text{ W/cm$^{2}$}$ with enlarged laser focal areas or high-$Z$ gases. These findings pave the way for the control of novel experiments such as the generation of pure electron-positron-photon plasmas from laser energy, and for probing QED in the extreme-intensity regime where the quantum vacuum becomes unstable.
\end{abstract}

\begin{document}

\flushbottom
\maketitle

\thispagestyle{empty}


\section*{Introduction}

As a consequence of the dramatic progress in high-power laser technology 
intensities of the order of $10^{24}\text{ W/cm$^2$}$ may become accessible 
in the next few years~\cite{eliURL, xcelsURL}, opening up the investigation of 
unexplored regimes of laser-matter interaction~\cite{marklundRMP06, mourouRMP06, 
ehlotzkyRPP09, dipiazzaRMP12, narozhnyCP15}. In particular, laser-induced pair production~\cite{ehlotzkyRPP09, dipiazzaRMP12, titovPRL12, suPRL12, blaschkePRD13, lvPRL13, dunneEPJST14, akalPRD14, 
wollertPRD15} and the development of electron-positron~($e^{-}e^{+}$) cascades~\cite{bellPRL08, 
kirkPPCF09, fedotovPRL10, elkinaPRSTAB11, nerushPRL11, duclousPPCF11, bulanovjrPRL10-1, gonoskovPRL13, kingPRA13, bashmakovPoP14, mironovPLA14, gelferPRA15, 
gonoskovPRE15, jirkaPRE16, grismayerPoP16, narozhnyPLA04, bulanovjrJETP06, fedotovLP09, 
ridgersPRL12} have attracted growing interest over the last decade. 
The main reason for this is their importance for providing a viable route to reproduce the 
extreme  conditions present in astrophysical environments~\cite{wardleN1998, remingtonRMP06} and shed light on the interplay between strong-field QED processes and 
multi-particle dynamics~\cite{dipiazzaRMP12, narozhnyCP15}.

Copious $e^{-}e^{+}\gamma$ production in the collision of two laser pulses may be 
initiated either by the initial presence of seed particles in the focal volume 
(seeded QED cascades), or by the spontaneous creation of $e^{-}e^{+}$ pairs out 
of the quantum vacuum (self-seeded QED cascades) when the laser field invariant 
$\mathcal{E}=\sqrt{(\mathcal{F}^2+\mathcal{G}^2)^{1/2}+\mathcal{F}}$ approaches 
the QED critical field $F_{\text{cr}} = m_e^2 c^3 / |e| \hbar \approx 1.3\times10^{16}\text{ V/cm}$, where $m_e$ and $e$ are the electron mass and charge, $c$ is the speed of light 
in vacuum, and $\hbar$ is the reduced Planck constant. Here, 
$\mathcal{F}\equiv(\mathbf{E}^2-\mathbf{B}^2)/2$ and $\mathcal{G}\equiv\mathbf{E}\cdot\mathbf{B}$ are the electromagnetic field invariants, with $\mathbf{E}$ and $\mathbf{B}$ being the electric and magnetic field, respectively. Seeded QED cascades occur because seed electrons are violently accelerated by the laser fields therefore emitting large amounts of hard photons which, in turn, convert into $e^{-}e^{+}$ pairs~\cite{bellPRL08, fedotovPRL10, elkinaPRSTAB11, nerushPRL11} via the multi-photon Breit-Wheeler process ($\gamma+n\gamma'\rightarrow e^-e^+$). The generated pairs are then accelerated by the laser fields and originate a new generation of particles. Such avalanche process was predicted to develop in the collision of two laser pulses with intensities around~\cite{bellPRL08} $10^{24}\text{ W/cm$^{2}$}$. By contrast, self-seeded cascades become significant for laser intensities beyond~\cite{bulanovjrJETP06} $10^{26}\text{ W/cm$^{2}$}$. Since the development of a seeded cascade initiated by stray electrons in imperfect vacuum would deplete the laser pulses already at intensities around~\cite{bellPRL08, fedotovPRL10, nerushPRL11} $10^{24}\text{ W/cm$^{2}$}$, it was pointed out that laser intensities approaching $10^{26}\text{ W/cm$^{2}$}$ might result inaccessible~\cite{bellPRL08, fedotovPRL10, nerushPRL11}. Thus, the determination of a mechanism to control the onset of seeded QED cascades is essential at ultra-high laser intensities. 

Hitherto, the theoretical analysis has been focused on the intensity required to trigger 
QED cascades whereas the implications of the strong field gradients associated with 
tightly focused laser pulses, which are essential to attain the highest intensities, have been neglected. Here we show the crucial role played by the laser field structure on 
the onset of seeded QED cascades when accounting for realistic laser pulse set-ups. 
On the one hand, tight laser focusing close to the diffraction limit may prevent 
the generation of seeded cascades up to intensities of the order of $10^{26}\text{ W/cm$^2$}$
per laser pulse. On the other hand, larger focal radii corresponding to five-six wavelengths 
allow for the onset of cascades already at much lower intensities below $10^{24}\text{ W/cm$^2$}$. Such a seemingly counter-intuitive trend with respect to the laser intensity is explained as tightly focused laser pulses, in addition to the limited available volume, necessarily exhibit strong gradients of the electromagnetic fields. These, in turn, result in large ponderomotive effects~\cite{mulser-book} with the expulsion of all stray electrons 
from the focal region occurring much before the peaks of the laser pulses field reach the focus. 
Hence, no stray electrons are available for initiating a cascade when the fields are close to 
their maximum even in a tenuous gas. Indeed, the generation of hard photons and $e^{-}e^{+}$ 
pairs may be completely suppressed. Note that QED processes, due to their stochastic nature, may also be suppressed in the interaction with very short laser pulses~\cite{harveyPRL17}.

In the following we investigate the conditions for the onset of $e^{-}e^{+}\gamma$ cascades initiated by stray electrons in the presence of two counter-propagating laser pulses both for aligned linear polarization (LP$_\parallel$), i.e. the electric fields of the two pulses are parallel to each other, and for crossed linear polarization (LP$_\perp$), i.e. the electric fields of the two pulses orthogonal to each other, depending on the waist radius $w_0$ and either the intensity $I = c E_0^2 / 8 \pi$, with $E_0$ being the peak field, or the power $P\approx\pi w_{0}^{2}I/2$ of each beam. QED cascades are characterized by an exponential growth in the number of $e^{-}e^{+}$ pairs with the creation on average of at least one pair per initial electron~\cite{fedotovPRL10, gelferPRA15}. Other scenarios where $e^{-}e^{+}$ pairs are produced but the above-mentioned conditions are not fulfilled are termed ``$e^{-}e^{+}\gamma$ gases''. In fact, in this case the electron and positron density remains small compared to the critical plasma density $n_{\text{cr}} \equiv m_e \omega^2 / 4 \pi e^2$, where $\omega$ is the laser frequency, and plasma effects are negligible over time scales comparable with the laser pulse duration.
In general, owing to the absence of a regular dynamics and of a guiding centre~\cite{bauerPRL95, kirkPPCF16}, the secular motion of electrons in a standing wave formed by two super-intense counter-propagating laser pulses may be investigated \emph{quantitatively} only by following the evolution of a statistical ensemble of particles~\cite{bauerPRL95, kirkPPCF16}. Hence, the equations of motion of a statistical ensemble of seed electrons were integrated numerically, and stochastic photon emission and pair production was modelled with a Monte Carlo approach which takes into account the effect of all photon emissions (see Supplementary Information for further details).

\section*{Results}

In our simulations the laser pulses propagate along the $z$ axis and their focus is located at the origin. \label{laser} They have Gaussian transverse spatial profile, $\lambda=0.8\text{ $\mu$m}$ wavelength, $T\approx 2.67\text{ fs}$ period, and sech$^2$ temporal intensity profile with $\tau\approx20\text{ fs}$ full-width at half maximum (FWHM) duration. The initial position $z_0$ of the laser peak intensity is at $z_0=\pm70\text{ }\lambda$. Initially, $10^3$ seed electrons are located at rest within a $\lambda^3$ volume at the laser pulse focus with uniform random distribution. The corresponding initial electron density $n_{e^-}(0)\approx 2\times 10^{15}\text{ cm$^{-3}$}$ is about $10^{-6}$ times the critical plasma density $n_{\text{cr}} \approx 1.7\times10^{21}\text{ cm$^{-3}$}$, such that here collective plasma effects are negligible for the \emph{onset} of a QED cascade. Indeed, plasma effects become appreciable for temporal scales of the order of the plasma period $T_{\text{pl}}=2\pi/\omega_{\text{pl}}$, where $\omega_{\text{pl}}=\sqrt{4\pi e^2 n_e /m_e}$ is the plasma frequency.
Since $T_{\text{pl}} = T \sqrt{n_{\text{cr}}/n_{e^-}(0)}\approx10^3\,T$, here we have $T_{\text{pl}}\gg\tau\approx 7.5\,T$ and throughout the laser pulse interaction electrons and positrons behave like a gas under the influence of external fields. Note that for relativistically intense laser pulses plasma effects are further suppressed, because in this case the electron effective mass~\cite{dipiazzaRMP12, mulser-book} $m_e^*=m_e\sqrt{1+\xi^2/2}$ replaces $m_e$ in $\omega_{\text{pl}}$, where $\xi=|e \textbf{E}|/m_e \omega c$ is the normalized laser field amplitude~\cite{dipiazzaRMP12, mulser-book}. 
Similar results are obtained by increasing by eight times both the initial volume and the number of seed electrons. 

Note that in laboratory conditions seed electrons originate from the ionization of gases which are unavoidably present even in vacuum chambers~\cite{klopferV60, fremereyV99}, hydrogen dissolved in stainless steel being the most significant source of residual gas in current ultra-high vacuum technology~\cite{fremereyV99}. The typical elements of such gases (H, C, N, and O) are ionized at electric field amplitudes comparable with the over-the-barrier ionization field $E_{\text{ion}}$, which are much lower than those reached at the focus of the high-intensity fields of interest here. Thus, bound electrons go into the continuum in the wings of the laser pulses and well before the laser peaks reach the focus. For hydrogen-like ions with atomic number $Z$ and in the ground state a quantum mechanical calculation gives~\cite{mulser-book}:
\begin{equation} \label{ionization}
E_{\text{ion}} = \frac{(\sqrt{2}-1)Z^{3}}{2\sqrt{2}}\frac{m_e^{2}|e|^{5}}{\hbar^{4}}.
\end{equation}
For H (O$^{7+}$) equation~(\ref{ionization}) provides $E_{\text{ion}}^{\text{H}}\approx 7.5\times 10^8\text{ V/cm}$ ($E_{\text{ion}}^{\text{O$^{7+}$}}\approx 3.9\times 10^{11}\text{ V/cm}$). Hence, depending on the gas, electrons are stripped off atoms at different values of the laser fields at the focal spot, i.e. at different distances of the peaks of the two laser pulses from the focus. In order to account for this critical effect, in all our simulations each seed electron is free to evolve just after the electric field at the electron position has risen above either $E_{\text{ion}}^{\text{H}}$ for hydrogen or $E_{\text{ion}}^{\text{O$^{7+}$}}$ for oxygen, which are the two cases investigated below. Also, in order to ascertain the influence of a phase-shift all simulations were performed with different relative phase between the two pulses (see Methods). We mention that for ultra-relativistic ions Eq.~\ref{ionization} would still be applicable but it would provide the value of the ionization field strength in the rest frame of the ion. However, ionization takes place in the wings of the laser pulses where the laser fields are relatively weak ($\xi \lesssim 1$ for low-$Z$ atoms) and, since the charge to mass ratio of ions is at least $2\times10^3$ times smaller than the electron change to mass ratio, ions remain non-relativistic or weakly relativistic during the whole ionization process. In fact, $\xi = |e E| \lambda / 2\pi m_e c^2$ approximately corresponds to the work performed by the laser field on an electron over a laser wavelength in units of the electron rest energy, such that electrons become relativistic when $\xi\approx1$. Since the mass of ions is $m_i\gtrsim 2\times10^3\,m_e$, this implies that $\xi \gtrsim 2\times10^3$ would be necessary for ions to become relativistic. Given the relatively low densities of the ions, after the ionization stage ions have no further effect on a QED cascade.

Strong-field QED effects are controlled by the electron/photon quantum parameter~\cite{Baier-book, ritusJSLR85} $\chi_{e/\gamma}=\sqrt{|(F_{\mu\nu}p^{\nu}_{e/\gamma})^2|}/F_{\text{cr}}m_e c$, where $F_{\mu\nu}$ is the field tensor and $p^{\nu}_{e/\gamma}$ the electron/photon four-momentum. For $\chi_{e}\ll 1$ the typical energy of the emitted photons is $\varepsilon_\gamma\approx\chi_e \varepsilon_e$, where $\varepsilon_e$ is the electron energy. For $\chi_{\gamma}\ll 1$ the probability of photon conversion into an $e^{-}e^{+}$ pair is exponentially suppressed~\cite{Baier-book, ritusJSLR85} as $e^{-8/3\chi_\gamma}$. Hence, single photon emission recoil is important as $\chi_e\gtrsim 1$ and pair creation is significant only when $\chi_\gamma\gtrsim 1$. 
Initially, the electrons from ionized atoms have relativistic factor $\gamma = \varepsilon_e/m_e c \approx 1$ such that they attain $\gamma \gg 1$ only after being accelerated by the fields of the laser pulses. For two colliding laser pulses with normalized field amplitude $\xi$ and wavelength $\lambda=0.8\,\mu\text{m}$, the electron quantum parameter can be estimated as $\chi_{e} \approx 6 \times 10^{-6}\gamma\xi$, while the photon quantum parameter is $\chi_{\gamma} \lesssim \chi_{e}$. Since here $\gamma \approx \xi$, one gets $\chi_{e}\gtrsim 1$ for a threshold amplitude of the order of $\xi_{\text{thr}} \gtrsim 400$, which corresponds to a laser intensity $I_{\text{thr}} \gtrsim 3.5\times10^{23}\,\text{W/cm$^2$}$. Given that here hard photon emission and electron-positron pair production only occur for $\xi_{\text{thr}} \gtrsim 400$, the formation length $\ell_f$, which is related to the field amplitude by\cite{ritusJSLR85, Baier-book} $\ell_f \approx \lambda / \xi$ remains always much smaller than the laser field extension in the regions where a QED cascade develops, because $w_0\gtrsim\lambda\gg \ell_f$. Thus, for $\xi \gg 1$ the locally-constant field approximation holds~\cite{ritusJSLR85}, and the formation length $\ell_f$ of quantum processes has no appreciable effect on the onset or development of a QED cascade as it is much smaller that the other relevant scales. In addition, since $\xi_{\text{thr}} \gg 1$ in the regions where a QED cascade occurs, virtual channels, i.e. those where off shell particles are involved, remain negligible. In fact, the probability of each real channel scales as $\alpha \ell / \ell_f$, where $\alpha\approx1/137$ is the fine structure constant and $\ell \lesssim \lambda$ is the extension of the considered field region, while the probability of a real two-step process scales as $(\alpha \ell / \ell_f)^2$. By contrast, the probability of a virtual channel with the same final states as the real two-step process scales as $\alpha^2 \ell / \ell_f$. Thus, the probability of a real processes is much larger than the corresponding virtual process when $\ell \gg \ell_f$, i.e. when $\xi \gg 1$ consistently with more detailed analytical calculations\cite{ritusJSLR85, kingPRD13, kingPRA15}.

\begin{figure}[ht]
\centering
\includegraphics[width=\linewidth]{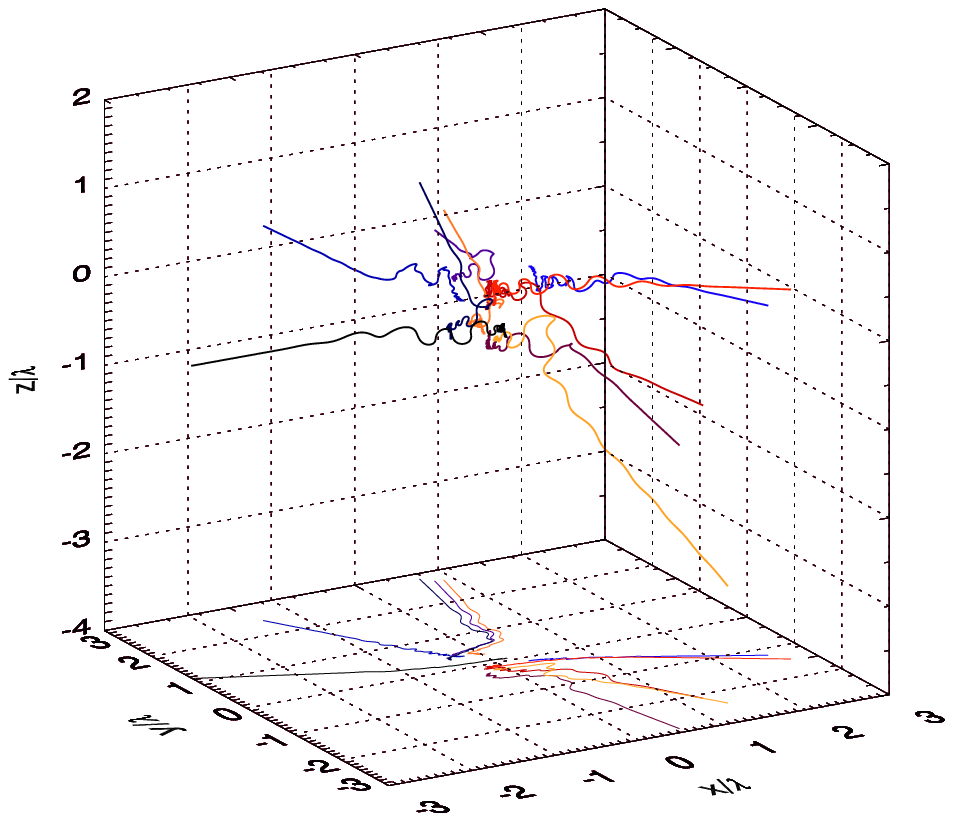}
\caption{\textbf{The trajectories of ten stray electrons driven by two tightly-focused counter-propagating laser pulses.} The displayed electron trajectories correspond to the temporal interval $0 < t < 36.5\text{ }T$, where $T=\lambda/c\approx2.67\text{ fs}$ and $\lambda=0.8\text{ $\mu$m}$ are the laser period and wavelength, respectively. Each laser pulse has $P=500\text{ PW}$ power and $w_0=1\text{ }\lambda$ waist radius, which implies an intensity $I\approx4.8\times10^{25}\text{ W/cm$^2$}$. The projection of the trajectories on the focal plane $xy$ is also reported.}
\label{trajectories}
\end{figure}

\begin{figure}[ht]
\centering
\includegraphics[width=\linewidth]{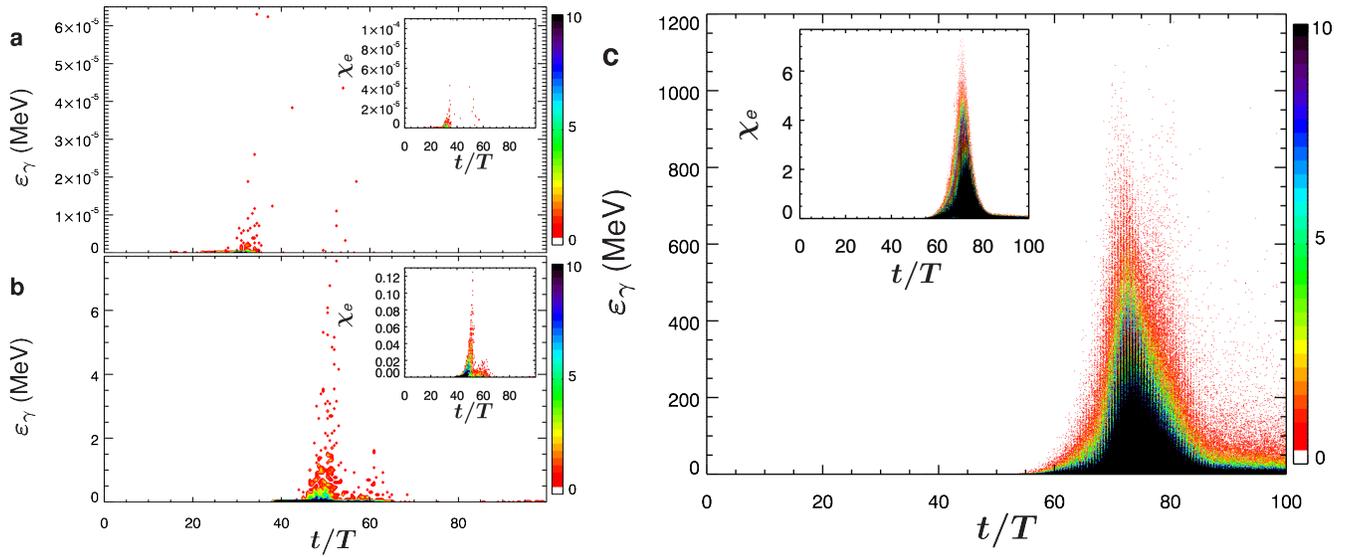}
\caption{\textbf{The distribution of the emitted photon energy as function of time.} \textbf{a}, The emitted photon energy $\varepsilon_\gamma$ and the electron quantum parameter $\chi_e$ (inset) at each photon emission event as function of time $t$. Time and space are in units of the laser period $T=\lambda/c\approx2.67\text{ fs}$ and laser wavelength $\lambda=0.8\text{ $\mu$m}$, respectively. Each laser pulse has $P=500\text{ PW}$ and $w_0=1\text{ }\lambda$, i.e. intensity $I\approx4.8\times10^{25}\text{ W/cm$^2$}$. Initially, the peaks of the two laser pulses are located at $z_0=\pm70\text{ }\lambda$. The colour bar levels correspond to the number of events (black indicates at least 10~events). Here stray electrons originate from hydrogen. \textbf{b}, Same parameters as in \textbf{a}, but stray electrons originate from oxygen. \textbf{c}, Same parameters as in \textbf{a}, but waist radius $w_0=7\text{ }\lambda$, i.e. intensity $I\approx 10^{24}\text{ W/cm$^2$}$.}
\label{phenergy}
\end{figure}

\begin{figure}[ht]
\centering
\includegraphics[width=\linewidth]{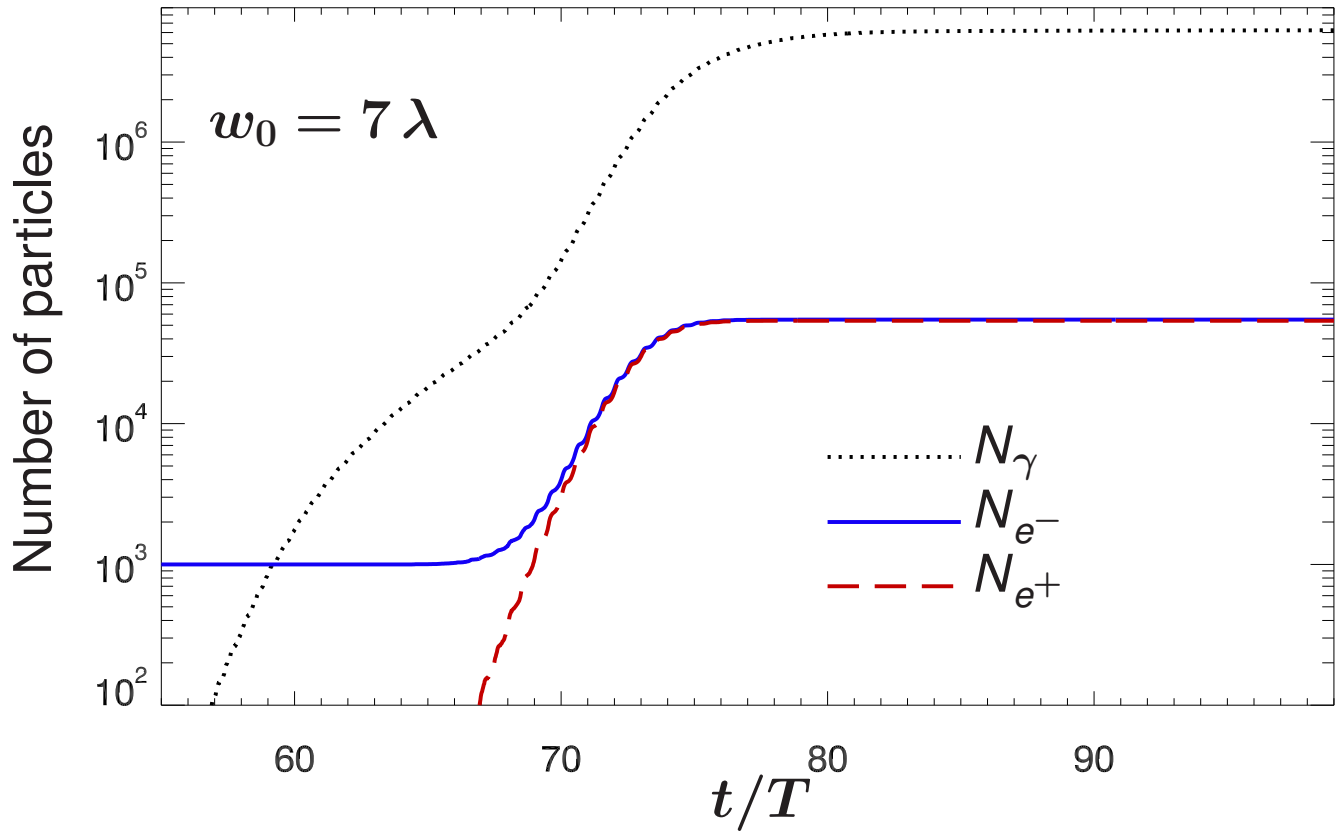}
\caption{\textbf{The evolution of the number of particles.} The total number of electrons $N_{e^-}$ (solid blue line), positrons $N_{e^+}$ (dashed red line) and photons $N_\gamma$ with energy $\varepsilon_\gamma > 2.5\text{ MeV}$ (dotted black line) as function of time $t$ during the collision of two counter-propagating laser pulses. Time and space are in units of the laser period $T=\lambda/c\approx2.67\text{ fs}$ and laser wavelength $\lambda=0.8\text{ $\mu$m}$, respectively. The two pulses have $P=500\text{ PW}$ power and $w_0=7\text{ }\lambda$ waist radius, which imply an intensity $I\approx 10^{24}\text{ W/cm$^2$}$. The initial $10^3$ stray electrons originate from hydrogen.}
\label{numberofparticles}
\end{figure}

Figure~\ref{trajectories} shows the trajectories of ten stray electrons originating from hydrogen in the interaction with two colliding laser pulses with LP$_\parallel$ polarization, $P$=500~PW and $w_0=1\text{ }\lambda$, i.e. $I\approx4.8\times10^{25}\text{ W/cm$^2$}$ per pulse. Electrons perform a complex motion inside the focal volume before being expelled by the strong field gradients~\cite{mulser-book}. However, all electrons have escaped the focal region at $t=36.5\text{ }T$, while the peaks of the two laser pulses are still at $z=\pm 33.5\text{ }\lambda$ from the origin, which implies an intensity of $2.8\times 10^{19}\text{ W/cm$^2$}$ at the focus. A similar result holds if the simulation is started with oxygen, except that inner shell electrons may remain bound until the laser pulse peaks are significantly closer to the focus. Hence, some electrons may remain in the focal volume up to $t\approx47\text{ }T$, which implies an intensity of $3.9\times 10^{21}\text{ W/cm$^2$}$ at the focus. Figures~\ref{phenergy}a and \ref{phenergy}b show the time dependence of $\varepsilon_\gamma$ and of $\chi_e$ at each emission event for the two above-mentioned cases. Since the laser fields at the focus are still relatively weak before the electrons exit, $\chi_e\ll 1$ and only low-energy photons $\varepsilon_\gamma\ll mc^2$ are emitted such that $e^{-}e^{+}$ pair production cannot occur. On the contrary, for $w_0\gtrsim 4\text{-}5\text{ }\lambda$ stray electrons are still inside the focal volume when the laser fields reach their maximum. Figure~\ref{phenergy}c displays $\varepsilon_\gamma$ and $\chi_e$ as functions of time with the same parameters as in Fig.~\ref{phenergy}a but $w_0=7\text{ }\lambda$. Although the laser pulse intensity decreases from $I\approx4.8\times10^{25}\text{ W/cm$^2$}$ to $I\approx10^{24}\text{ W/cm$^2$}$, in this case $\chi_e>1$ and, in sharp contrast with the results reported in Fig.~\ref{phenergy}a, copious emission of hard-photons with several hundreds MeV energy occurs. Figure~\ref{numberofparticles} shows the evolution of the total number of photons $N_\gamma$ with $\varepsilon_\gamma>2.5\text{ MeV}$, the number of electrons $N_{e^-}$ and positrons $N_{e^+}$ as functions of time. As $N_\gamma$ and the intensity at the focus rise, a significant fraction of the emitted high-energy photons convert into $e^{-}e^{+}$ pairs therefore initiating an $e^{-}e^{+}\gamma$ cascade with exponential increase of the number of pairs. The total number of emitted photons is about $1.44 \times 10^7$, of which about $6.3 \times 10^6$ with $\varepsilon_\gamma>\text{ 2.5~MeV}$.  Also, approximately $5.4\times 10^4$ $e^{-}e^{+}$ pairs were produced, which confirms that soft-energy photons, despite of their large number, are not capable of preventing the onset or the development of $e^{-}e^{+}$ cascades. Finally, for $w_0\approx3\;\lambda$ transition regions exist, where the onset of a QED cascade is sensitive to the precise initial conditions such as the relative phase of the laser pulses.

Note that Figs.~\ref{phenergy}c,~\ref{numberofparticles} display the results obtained with the same laser power as in Figs.~\ref{phenergy}a,~b but larger $w_0$. 
For $w_0=1\text{ }\lambda$ no seed electron is present to initiate a cascade (see Fig.~\ref{phenergy}a), while for $w_0=7\text{ }\lambda$ prolific pair production is observed (see Figs.~\ref{phenergy}c,~\ref{numberofparticles}). Following Ref.~\cite{bashmakovPoP14}, the final positron yield $N_{e^{+}}(\infty)$ can be parametrized as:
\begin{equation} \label{rate}
N_{e^{+}}(\infty) = \frac{N_{e^{-}}(0)}{2}\left[e^{\int_{0}^{\infty}{\Gamma(t)dt}}-1\right] \approx \frac{N_{e^{-}}(0)}{2} [e^{\langle\Gamma\rangle \tau}-1],
\end{equation}
where $\Gamma(t)$ and  $\langle\Gamma\rangle$ are the instantaneous growth rate and its average over $\tau$ around the peak laser intensity, respectively ($\langle\Gamma\rangle\approx0.6\text{ }T^{-1}$ and $n_{e^\pm}(\infty)\approx54\text{ }n_{e^-}(0)\lesssim 10^{-4} n_{\text{cr}}$ in Fig.~\ref{numberofparticles}).
For fixed laser power, the larger $\langle\Gamma\rangle$ corresponds to the higher laser intensity where no seed electron expulsion occurs before the laser peaks reach the focus (see Table~\ref{Tab1}). 
As $w_0$ is increased to larger values, $\langle\Gamma\rangle$ gradually decreases until almost no pair is produced because the laser intensity becomes too low. 

\section*{Discussion}

The results of our simulations are collected in Table~\ref{Tab1}. Table~\ref{Tab1} shows that:
\begin{enumerate}
\item for $w_0\lesssim 1\text{-}2\text{ }\lambda$ no $e^{-}e^{+}$ pairs are produced even at an intensity of $10^{26}\text{ W/cm$^2$}$,
\item $e^{-}e^{+}$ pair creation is more suppressed for LP$_\perp$ than for LP$_\parallel$, and
\item the nature of the gas is of critical importance for determining the minimal power and intensity required to initiate seeded QED cascades.
\end{enumerate}
In fact, if suitable high-$Z$ gases such as krypton or xenon are introduced into the vacuum chamber, their inner shell electrons may remain bound during the rise of the laser pulses and go into the continuum only as the laser fields at the focus are close to their maximum~\cite{huPRL02, hetzheimPRL09}. For example, our simulations show that if electrons go into the continuum when the peaks of the pulses are at $z_0=\pm3\text{ }\lambda$, the required laser power falls from approximately 200~PW to 11~PW per pulse with $w_0=0.9\text{ }\lambda$ and $I\approx1.3\times10^{24}\text{ W/cm$^2$}$.

The above results have been obtained with a sech$^2$ temporal intensity profile
because its shape bears a closer similarity to those obtained in experiments compared 
to the more widely used Gaussian temporal profile~\cite{asakiOL93, lazaridisOL95}. 
In fact, although sech$^2$ and Gaussian profiles look quite similar, only the 
sech$^2$ profile presents exponential rise and fall wings, which corresponds to 
the prediction of the laser rate equations (see Refs.~\cite{asakiOL93, lazaridisOL95} and 
references therein). In addition, for a sech$^2$ temporal intensity profile an accurate description of the fields of a tightly focused laser pulse is available for its whole temporal domain (see Methods for details). This is important because both gas ionization and electron expulsion occur mainly in the wings of the laser pulse and well before the peaks of the laser pulses reach the focus. Thus, for accurate \emph{quantitative} predictions, the laser pulse structure should be known both in the temporal~\cite{asakiOL93, lazaridisOL95} and in the transverse domain\cite{patelPPCF05}. Notice that, unlike in the case of a Lorentzian transverse profile\cite{patelPPCF05}, in the case of a transverse Gaussian profile fully analytical expressions of the electromagnetic field structure beyond the paraxial approximation are available. This is essential when tightly focused laser pulses are considered~\cite{salaminPRL02}.

We stress that, although ponderomotive particle expulsion from the focal volume due to the strong field gradients of tightly focused laser pulses is a \emph{general} phenomenon which is known to occur also for a single laser pulse\cite{mulser-book}, already in the relatively simple case of two colliding plane waves the electron dynamics may become sensitive to the precise initial conditions~\cite{bauerPRL95}. Thus, it is not possible to provide a reliable and quantitative \emph{analytical} prediction of the parameters required to cut off a QED cascade, which depends on the temporal and transverse shape of the laser pulses, including the laser pulse duration, polarization and the orientation of the polarization axis, stochastic effects intrinsic in both ionization and QED processes and, at the threshold, even on the relative phases of the laser pulses (see Table~\ref{Tab1}). This implies that for \emph{quantitative} predictions one has to resort to numerical simulations of statistical ensembles of particles. Indeed, simple analytical models of the electron dynamics as those considered e.g. in Ref.~\cite{fedotovPRL10}, although are useful for a qualitative understanding of the problem, cannot account for the actual complex electron dynamics. Thus, as it is indicated by the same authors in Ref.~\cite{fedotovJPCS16}, they are unable to predict analytically, for example, the time an electron needs to escape the laser focus, which is of essential importance for determining the parameters required for the onset of a QED cascade. We mention that, when a QED cascade develops, the time an electron needs to escape the laser focus is also of critical importance for determining the number of cascade generations produced. In fact, the number of  cascade generations depends directly on the ratio between the escape time and the mean time required for photon emission and for pair-production. 
For example, with the same parameters as on page~\pageref{laser} but $\tau\approx10\text{ fs}$, for LP$_{\parallel}$ (LP$_{\perp}$) and $I\gtrsim5\times10^{24}\text{ W/cm$^2$}$ a seeded QED cascade arises already for $w_0\gtrsim2\text{ }\lambda$ ($w_0\gtrsim3\text{ }\lambda$) with hydrogen, while a transition region where cascades may develop exists already for $w_0\approx1\text{ }\lambda$ ($w_0\approx2\text{ }\lambda$) with oxygen. 

Note that seeded QED cascade suppression was also obtained in a planar set-up with four tightly focused laser pulses colliding at $90^\circ$ with 250~PW power per pulse, aligned polarization axis, $w_0=1\text{ }\lambda$, and the other parameters as on page~\pageref{laser}. In fact, albeit seed electrons expulsion from the focus diminishes as compared to the case of two pulses, only in very special cases $e^{-}e^{+}$ pair production may occur: if the initial phases are chosen such that almost perfect cancellation of the electric field at the focus occurs (both for hydrogen and for oxygen), or if the system is perfectly symmetric with the four pulses exhibiting also almost the same initial phase (only for oxygen). Indeed, by performing $10^3$ simulations with four identical laser pulses but randomly chosen initial laser pulse phases and hydrogen (oxygen) residual gas, only for 2.6\% (38.6\%) of the simulations a QED cascade was initiated. Moreover, the occurrence of the above-mentioned special circumstances can be avoided for arbitrary initial phases by employing linearly polarized laser pulses with mutually oblique polarization axis. Thus, the structure of the laser pulses and the nature of the gas are of critical importance for the onset of seeded QED cascades also in more complex set-ups with multiple colliding laser pulses.

In summary, the general key message of this work is that, in addition to the laser intensity, the onset of QED cascades depends critically also on the laser pulse structure and on the nature of the gas. Moreover, the onset of QED cascades can be controlled via the laser pulse waist radius or by changing the atomic species of the residual gas.

\section*{Methods}

\textbf{Numerical Modelling.} All our simulations were performed employing both a standard Boris pusher~\cite{birdsall-langdon} with a time-step much smaller than the laser period $\Delta t = 1 / (10^{5} \omega)$ where  $\omega = 2 \pi / T$, and an adaptive fourth order Runge-Kutta integrator with time step chosen such that $\Delta t < T / (10 \xi_M)$, where $\xi_M = e E_M / m_e \omega c$ is the normalized field amplitude and $E_M$ is the maximum of the local value of the electric and magnetic field at the particle's position. For both integrators, stochastic photon emission by electrons and positrons and electron-positron pair creation from high-energy photons was taken into account by employing a standard Monte Carlo technique~\cite{elkinaPRSTAB11}. The small time step renders negligible the probability of multiple photon emission events during each time step. No significant difference was found between the two different integrators. Further details on the Monte Carlo technique, and benchmarks of the code against published results on the formation of a seeded QED cascade are reported in the Supplementary Information. For photons with energy less than 2.5~MeV and in the regions where $\chi_\gamma<0.3$ photon conversion into pairs was neglected because for these photons the mean free path for pair conversion is much longer than the considered laser pulse duration.

In all our simulations a fully three-dimensional description of the laser pulse fields with terms up to the fifth order in the diffraction angle $\epsilon = \lambda / \pi w_0$ was employed~\cite{salaminAPB07}. Also, all simulations were performed with zero, $\pi/2$ and $\pi$ relative phase between the two pulses. For LP$_\parallel$ the three relative phases imply that the superposition of the laser fields at the focus has vanishing, non vanishing and maximal magnetic field component, respectively. Note that, in general, there exist spatio-temporal regions where a description of the laser pulse fields as the product of a temporal and of a transverse beam profile does \emph{not} provide a good approximate solution of Maxwell equations~\cite{mcdonaldURL}. In fact, for a temporal field envelope $g(\varphi)$ where $\varphi$ is the laser pulse phase, terms that mix the temporal and transverse profile of the laser pulse are important in the regions where $|g'/g| \gtrsim 1$, with $g'\equiv dg(\varphi)/d\varphi$. For a Gaussian temporal profile $|g'/g|\gtrsim1$ in the wings of the pulse~\cite{mcdonaldURL}, which is the region where ionization and electron expulsion mainly occurs. By contrast, for the sech (sech$^2$) temporal field (intensity) profile $g(\varphi)=\text{sech}(\varphi/\varphi_0)$ one obtains $|g'/g|=\text{tanh}(\varphi/\varphi_0)/\varphi_0<1/\varphi_0$, which remains much smaller than the unity in the whole temporal domain for $\varphi_0\gg1$ (see Ref.~\cite{mcdonaldURL} for further details). For a sech$^2$ laser pulse with $\tau\approx20$~fs duration FWHM of the intensity one can estimate a better than 3.7\% accuracy in the whole temporal domain.
In our simulations the laser fields were considered as a given classical background because collective plasma effects, such as laser pulses depletion or self-generated fields, were completely negligible for the onset and initial development of a QED cascade. 
The initial position $z_0$ of the two laser peaks was chosen at $z_0=\pm70\text{ }\lambda$ to ensure the the laser fields at the focus are initially much smaller than the over-the-barrier ionization field of the considered residual gases. 

The simulation results shown in Figs.~\ref{trajectories},~\ref{phenergy}, and~\ref{numberofparticles} were obtained by starting the simulation with LP$_\parallel$ polarization and zero relative phase between the two pulses. No qualitative difference is found by changing the relative phase of the two pulses.

\textbf{Data Availability.} The datasets generated and analysed during the current study are available from the corresponding author on reasonable request.



%

\section*{Author contributions statement}

M.T. initially conceived the project, wrote the numerical code, carried out the simulations, analysed the results and wrote the bulk of the manuscript. A.D.P. provided theoretical support for the analysis of the simulations and the interpretation of the results. C.H.K provided overall supervision. All authors contributed to the preparation of the manuscript.

\section*{Additional information}

\textbf{Competing financial interests} 
The authors declare no competing financial interests.

\begin{table}[ht]
\centering
\begin{tabular}{|l c c c c | c c c|}
\hline
$I$~(W/cm$^2$) & $w_0$~$(\lambda)$ & \multicolumn{3}{c}{LP$_{\parallel}$ (\textbf{H} $|$ \textit{O})} & \multicolumn{3}{c}{LP$_{\perp}$ (\textbf{H} $|$ \textit{O})} \\
\hline
\multirow{6}{*}{$10^{24}$} & 1 & \textbf{N}&$|$&\textit{N} & \textbf{N}&$|$&\textit{N} \\
  & 2 & \textbf{N}&$|$&\textit{G} & \textbf{N}&$|$&\textit{N} \\
  & 3 & \textbf{N$\leftrightarrow$G}&$|$&\textit{G$\leftrightarrow$C} & \textbf{N}&$|$&\textit{N$\leftrightarrow$G} \\
  & 4 & \textbf{N$\leftrightarrow$G}&$|$&\textit{C} & \textbf{N$\leftrightarrow$G}&$|$&\textit{G} \\
  & 5 & \textbf{G$\leftrightarrow$C}&$|$&\textit{C} & \textbf{G}&$|$&\textit{G} \\
  & $\geq 6$ & \textbf{C}&$|$&\textit{C} & \textbf{C}&$|$&\textit{C} \\
\hline
\multirow{5}{*}{$10^{25}\text{ \& }10^{26}$} & 1 & \textbf{N}&$|$&\textit{N} & \textbf{N}&$|$&\textit{N} \\
  & 2 & \textbf{N}&$|$&\textit{C} & \textbf{N}&$|$&\textit{N} \\
  & 3 & \textbf{N$\leftrightarrow$C}&$|$&\textit{C} & \textbf{N}&$|$&\textit{N$\leftrightarrow$C} \\
  & 4 & \textbf{C}&$|$&\textit{C} & \textbf{N$\leftrightarrow$C}&$|$&\textit{C} \\
  & $\geq 5$ & \textbf{C}&$|$&\textit{C} & \textbf{C}&$|$&\textit{C} \\
\hline\hline
$P$~(PW) & $w_0$~$(\lambda)$ & \multicolumn{3}{c}{LP$_{\parallel}$ (\textbf{H} $|$ \textit{O})} & \multicolumn{3}{c}{LP$_{\perp}$ (\textbf{H} $|$ \textit{O})} \\
\hline
\multirow{6}{*}{200} & 1 & \textbf{N}&$|$&\textit{N} & \textbf{N}&$|$&\textit{N} \\
  & 2 & \textbf{N}&$|$&\textit{C$_{2.4}$} & \textbf{N}&$|$&\textit{N} \\
  & 3 & \textbf{N}&$|$&\textit{C$_{1.4}$} & \textbf{N}&$|$&\textit{N$\leftrightarrow$C} \\
  & 4 & \textbf{G$\leftrightarrow$C}&$|$&\textit{C$_{0.7}$} & \textbf{N}&$|$&\textit{G$\leftrightarrow$C} \\
  & 5 & \textbf{G}&$|$&\textit{C$_{0.3}$} & \textbf{G}&$|$&\textit{G} \\
  & $[6, 9]$ & \textbf{G}&$|$&\textit{G} & \textbf{G}&$|$&\textit{G} \\
\hline
\multirow{7}{*}{500} & 1 & \textbf{N}&$|$&\textit{N} & \textbf{N}&$|$&\textit{N} \\
  & 2 & \textbf{N}&$|$&\textit{C$_{6.0}$} & \textbf{N}&$|$&\textit{N} \\
  & 3 & \textbf{N$\leftrightarrow$C}&$|$&\textit{C$_{3.7}$} & \textbf{N}&$|$&\textit{N$\leftrightarrow$C} \\
  & 4 & \textbf{C$_{1.6}$}&$|$&\textit{C$_{2.4}$} & \textbf{N}&$|$&\textit{C$_{2.1}$} \\
  & $[5, 8]$ & \textbf{C$_{[1.3,\,0.3]}$}&$|$&\textit{C$_{[1.7,\,0.4]}$} & \textbf{C$_{[0.7,\,0.3]}$}&$|$&\textit{C$_{[1.1,\,0.2]}$} \\
  & 9 & \textbf{C$_{0.2}$}&$|$&\textit{C$_{0.2}$} & \textbf{G$\leftrightarrow$C}&$|$&\textit{G$\leftrightarrow$C} \\
  & $[10, 15]$ & \textbf{G}&$|$&\textit{G} & \textbf{G}&$|$&\textit{G} \\
\hline
\end{tabular}
\caption{\label{Tab1}\textbf{Seeded QED cascade regimes.} 
Results are reported for aligned (LP$_{\parallel}$) and orthogonal (LP$_{\perp}$) linear polarization and either for hydrogen (H, bold left) or for oxygen (O, italic right). Symbols are as follows: N = No $e^{-}e^{+}$ pairs, G = $e^{-}e^{+}$ gas, C = $e^{-}e^{+}\gamma$ cascade, A$\leftrightarrow$B = transition between regime A and B. The subscripts indicate the average cascade growth rate $\langle\Gamma\rangle$ in units of $T^{-1}$.}
\end{table}


\end{document}